\documentclass[12pt,letterpaper]{article}
\usepackage[utf8]{inputenc}
\usepackage[version=3]{mhchem} 
\usepackage{siunitx} 
\usepackage{graphicx} 
\usepackage{natbib} 
\usepackage{amsmath} 
\usepackage{graphicx}
\usepackage[outdir=./]{epstopdf}
\usepackage{natbib}
\usepackage{multirow}
\usepackage{xcolor}
\usepackage{newtxtext,newtxmath}
\usepackage{graphicx,subfigure}
\usepackage{multirow}
\usepackage{tabulary,booktabs}
\usepackage{lineno}
\usepackage{amsmath}
\usepackage{hyperref}
\usepackage{xspace}
\usepackage{array}

\title{Complex Network View of the Sun's Magnetic Patches: I. Identification} 
\author{Zahra \textsc{Tajik}$^{1}$\thanks{E-mail: z.tajik@znu.ac.ir}, Nastaran \textsc{Frahang}$^{3}$\thanks{E-mail: farhangnastaran@gmail.com}, Hossein \textsc{Safari}$^{1,2}$\thanks{E-mail: safari@znu.ac.ir}\\
	, and Michael S. \textsc{Wheatland}$^{3}$\thanks{E-mail: michael.wheatland@sydney.edu.au}}

\date{$^{1}$Department of  Physics, Faculty of Science, University of Zanjan, University Blvd., Zanjan, 45371-38791, Zanjan, Iran\\%
$^{2}$Observatory, Faculty of Science, University of Zanjan, University Blvd., Zanjan, 45371-38791, Zanjan, Iran\\%
$^{3}$Sydney Institute for Astronomy, School of Physics, The University of Sydney, NSW 2006, Australia\\%
\today
} 

\begin{document}

\maketitle

\section{Abstract} 
Solar and stellar magnetic patches (i.e., magnetic fluxes that reach the surface from the interior) are believed to be the primary sources of a star's atmospheric conditions.
Hence, detecting and identifying these features (also known as magnetic elements) are among the essential topics in the community. Here, we apply the complex network approach to recognize the solar magnetic patches. For this purpose, we use the line-of-sight magnetograms provided by the Helioseismic and Magnetic Imager on board the Solar Dynamic Observatory. We construct the magnetic network following a specific visibility graph condition between pairs of pixels with opposite polarities and search for possible links between these regions. The complex network approach also provides the ability to rank the patches based on their connectivity (i.e., degree of nodes) and importance (i.e., PageRank). The use of the developed algorithm in the identification of magnetic patches is examined by tracking the features in consecutive frames, as well as making a comparison with the other approaches to identification. We find that this method could conveniently identify features regardless of their sizes. For small-scale (one or two pixels) features, we estimate the average of 8\% false-positive and 1\% false-negative errors.

\section{Introduction}
In Sun-like stars, the magnetic field is transferred from the inner layers to the atmosphere as the buoyantly unstable field lines bundle into the convection zone, stretch and twist along the path, and finally break through the surface. Such a dynamo creates a complex magnetic environment at which new fluxes (i.e., magnetic patches) continuously appear and cancel on a star's surface \citep{parker1955hydromagnetic, murray20063d, priest2014magnetohydrodynamics, schmieder2014magnetic, Farhang2018}. Accordingly, the solar photosphere is covered by magnetic features of various sizes and time scales ranging from tiny granular magnetic loops with fluxes as small as $10^{16}$ Mx and lifetimes of a few seconds/minutes to active regions (ARs) with fluxes up to $10^{23}$ Mx and typical lifetimes of several weeks \citep{zwaan1985emergence,hagenaar1999dispersal, wiehr2004brightness, cheung2007magnetic, tortosa2009magnetic, priest2014magnetohydrodynamics, archontis2019emergence}. 
It is important to note that no explicit definition has been introduced for magnetic features on the Sun's surface, yet, the term is commonly used in referring to flux concentrations and ephemeral regions \citep{deforest2007solar}.

\begin{figure*}
   \centering
	   \begin{center}$
	    \begin{array}{cc}
	        \includegraphics[width=7.4cm,height=4.9cm]{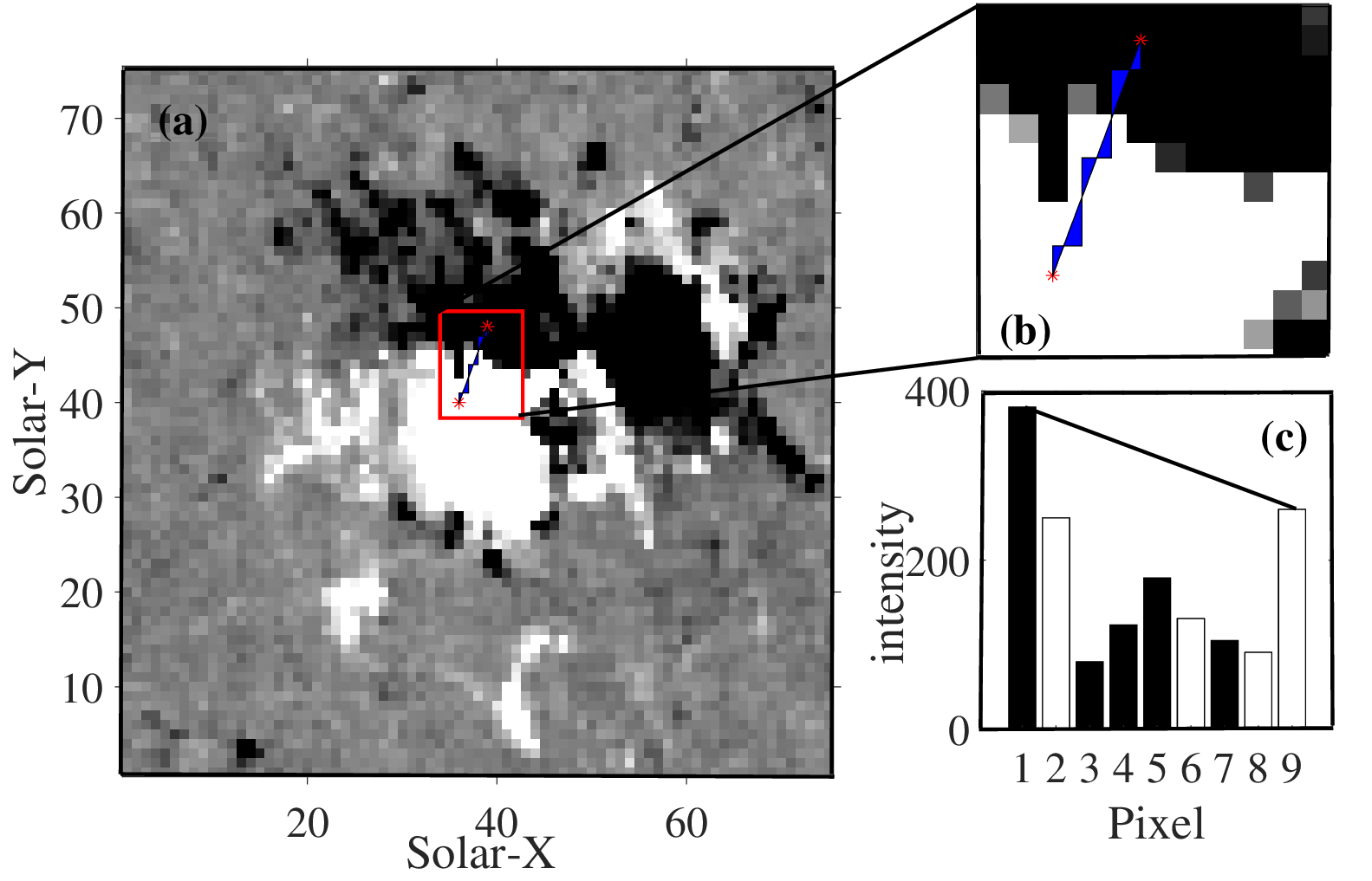}&
             \includegraphics[width=4.5cm,height=4.5cm, angle=1]{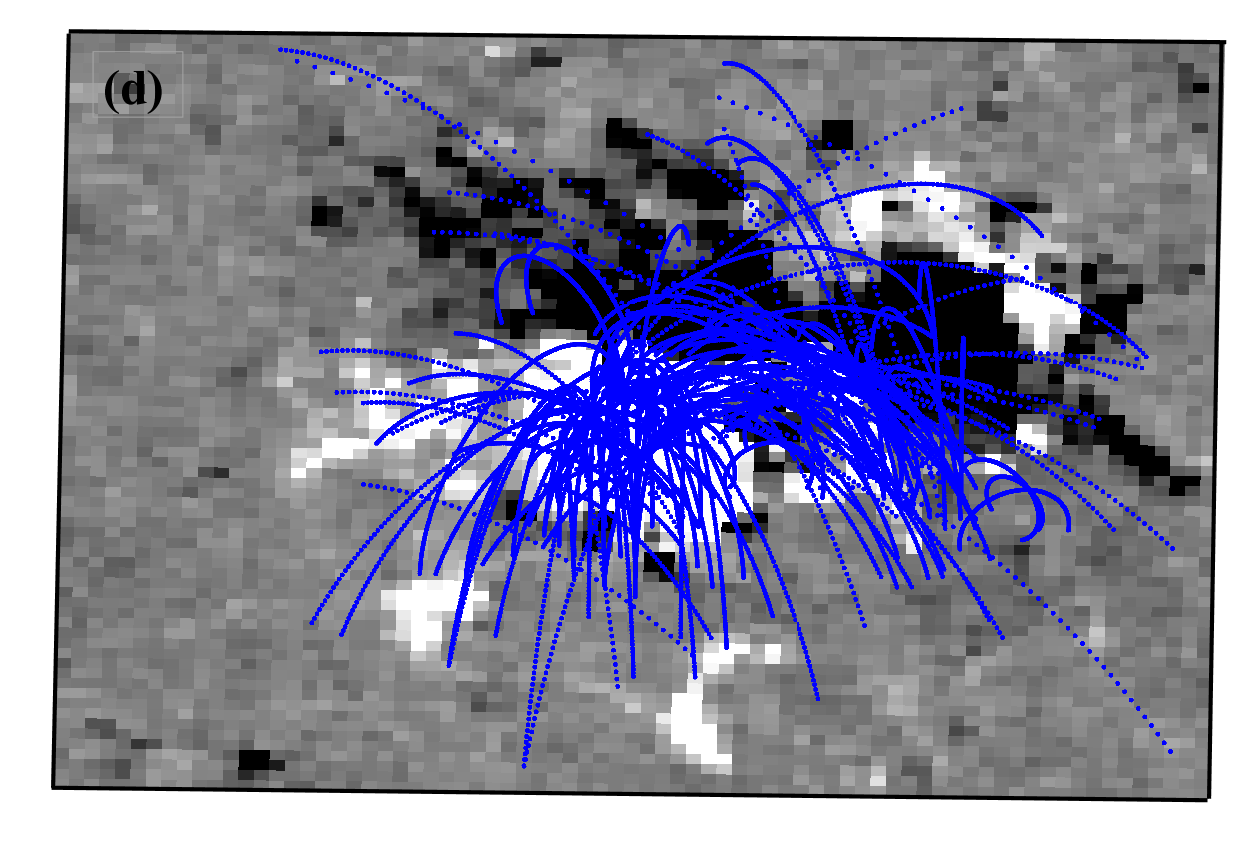}
          \end{array}$
         \end{center}
    \caption{An HMI cutout image of the Sun taken by the SDO at 23:58 on January 17, 2022. As manifested in this graphical illustration, an arbitrary pixel could connect an opposite polarity (red asterisks) only if their magnetic intensities exceed the absolute values of all the pixels placed in between them (marked with blue flags in panel (b)). The absolute intensities of negative and positive fluxes are shown with black and white bars in panel (c), respectively. A partial 3D visualization of the magnetic connections that hold in the constructed network is displayed in panel (d). The heights of connections represent their weights.}
   \label{Fig1}
\end{figure*}

To this date, extensive research has been devoted to the observation and recognition of solar magnetic patches and various routines have been developed aspiring this purpose. These algorithms could principally be classified into threshold-based, region-growing-based, and clustering-based segmentation methods, as well as deep learning algorithms \citep[e.g.,][]{Welsch2003, mcateer, benkha, deforest2007solar, barra, Watson, barrafast, zhang, harker, verbeeck, bo2022automatic}. 
\citep{hagenaar1999dispersal} used the curvature of the two-dimensional map of the Michelson Doppler Imager (MDI) values of $B_\textrm{LOS}$ at the photosphere to detect the small solar magnetic patches. The magnetic concentrations are then determined through a pixel-clumping algorithm. \citep{parnell2002nature} proposed the Magnetic Clumping Associative Tracker (MCAT) method which identifies magnetic features of the quiet Sun by applying a threshold-based technique. MCAT uses the intensity distribution of MDI images and implements a Gaussian approach \citep[also see,][]{lamb2003magnetic}. Yet Another Feature Tracking Algorithm (YAFTA) was designed in 2002 to detect both small and large-scale magnetic patches \citep{welsch2002magnetic}. Given an initial intensity threshold, YAFTA applies a gradient-based method, namely the Downhill method, to extract the positive/negative field concentrations from magnetogram images of MDI \citep{deforest2007solar}.

In early 2000, with the extensive development of artificial intelligence techniques, new generations of routines were developed to tackle the patch identification problem \citep{mcateer, zharkov2005statistical}. \citep{Colak} developed an algorithm using both the image processing techniques (i.e., the morphological procedure, watershed transform, image enhancement routine, and region-growing method) and the machine learning approach (i.e., the neural network) to discern the solar disk borders in H$\alpha$ images of the Sun, eliminate the limb-darkening effect, and track ARs. \citep{barra} introduced the Spatial Possibilistic Clustering Algorithm  that divides solar full-disk EUV images into coronal Holes, ARs, and quiet regions via the Fuzzy Compact Clustering Means (FCM), and Possibilistic Compact Clustering Means (PCM) algorithms. \citep{kestener} performed a wavelet-based analysis on the magnetogram images to detect solar ARs and studied the multi-fractal characteristics of these features. 

\citep{Higgins} investigated ARs via a region-growing perspective based on the magnetic field strength. \citep{caballero} introduced a three-step algorithm to identify magnetic patches from the EUV images of the Sun. In this method, the images are first segmented into regions with similar properties (according to their intensity histograms). Then, the segments are classified following a hierarchical procedure. Finally, the results are validated through an optimization problem. \citep{mohsen} developed an unsupervised segmentation routine based on the Bayesian approach to distinguish between solar ARs and coronal holes in EUV images. \citep{quan} employed a deep learning algorithm to determine boundaries of the photospheric fluxes that appeared on the mid-longitudes of the solar disk between 2010 and 2017. They applied a convolutional neural network as well as the YOLO-V3 algorithm and compared the efficiency of these methods. 

Despite all the advances made in the investigation of solar atmospheric patches over the past century, the true nature of their underlying mechanism is yet to be understood \citep{cho2007magnetic,bellot2019quiet, Farhang2019, Farhang2022}. The study of the photospheric flux concentrations and their dynamic evolution could provide new insight into the physical processes responsible for the generation and transport of the solar/stellar magnetic field \citep{deforest2007solar, kosovichev2009photospheric}. Furthermore, the detection and tracking of the surface fields might even deliver forecasting capabilities \citep{nobrega2020nonequilibrium}. Here, we apply the complex network approach and discuss the use of such a novel perspective in the recognition of flux concentrations. The temporal evolution of magnetic patches, as a crucial step in the identification of ARs, will be appraised within the next article in the series.

The complex network approach has recently become of interest for studying solar magnetic structures \citep{gheibi, Farhad, najafi,taran2022complex}. This method provides a powerful tool for the identification and examination of complex systems \citep{donges, boers, kaki}. Here, we use the magnetogram images to construct the solar magnetic network and assess its dynamic properties. According to the network theory, a graph (a set of edges and nodes as a mathematical representation of a network) could describe the complex relations governing a system \citep{Steinhaeuser, KARSTEN}. In such a schematic, nodes (vertices) may represent geographical regions, or any other concept depending on the subject system and an edge (link) indicates a physical or mathematical correspondence between two nodes. We consider the pixel locations as our nodes, and links are established if some predefined criteria are met. Generally, various types of graphs are possible depending on the intended condition (e.g., correlation-based, visibility-based, etc.) and the establishment of connections (i.e., simple, directed, or weighted). We show that the complex network approach could detect and identify magnetic features with great accuracy. 

The remainder of this paper is organized as follows: we introduce the employed data set in Section \ref{sec:data}. The details of the developed method and performed analysis are presented in Sections \ref{sec:network} and \ref{sec:properties}, respectively. The obtained results are discussed in Section \ref{sec:results}.

\section{Data}\label{data}
Solar magnetic patches have been observed over decades by various land-based and space-based instruments \citep[see e.g.,][and the references therein]{miesch2005large,bellot2019quiet}. The Solar Dynamics Observatory (SDO) mission, launched in 2010, is a well-equipped spacecraft that has provided high-quality data in recent years. One of the instruments on board the SDO is the Helioseismic and Magnetic Imager (HMI). This telescope is mainly designed to study the complex evolution of the solar magnetic field and its origin in both the inner and outer layers of the Sun \citep{scherrer}.

\begin{figure*}
 \centering
	   \includegraphics[width=12cm,height=3cm]{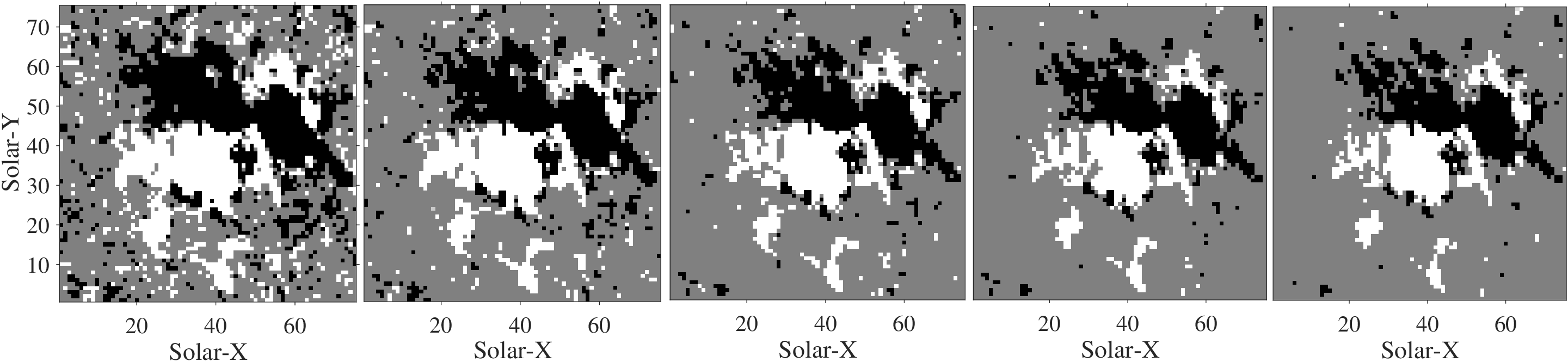}	
		\caption{The identified magnetic patches of the HMI image of Figure \ref{Fig3}(a). The background magnetic threshold is considered 8, 12, 16, 20, and 24 G, from left to right respectively.}
		\label{Fig2}
\end{figure*}

HMI provides full-disk images in the absorption line \textrm{Fe}$\mathrm{I}$  at $6173$ {\AA}, with the spatial resolution of $1^{\prime\prime}$ and temporal resolution of $45$ s \citep{schou2012design}. This instrument registers Dopplergram images (solar surface velocity maps), continuum filtergrams (wide spectrum images of the shadows), line-of-sight (LOS), and vector magnetograms (magnetic field maps of photosphere) \citep{Pesnell, DeRosa}. Data are available at Stanford University's Joint Science Operations Center (JSOC) database at \url{jsoc.stanford.edu}. The JSOC catalog archives the data in different resolutions of $4096 \times 4096$, $2048 \times 2048$, and $1024 \times 1024$ pixels. 

We use the HMI LOS magnetograms with the spatial sampling of $2.4^{\prime\prime}$ pixel$^{-1}$ at $1024 \times 1024$ pixels with $45$ s intervals. $B_\textrm{LOS}$ is the radial field component at the disk center but includes non-radial components away from the disk center. However, the LOS magnetic field variation is insignificant and neglected in cases with the typical field-of-view of an AR.

\section{The Magnetic Complex Network}\label{sec:network}

We aim to construct the magnetic complex network and evaluate its utility in the recognition of solar photospheric patches. Applying this approach could improve our understanding of the Sun's atmospheric events and their origins. The first step to establishing a network is defining nodes and edges. We consider each pixel of the HMI image as a node, and the existence of a link between each pair of nodes is verified based on the visibility graph condition:

\begin{equation}
    \label{eqVG}	
 I_{\textrm{i1,j1}}, I_{\textrm{i2,j2}} > I_{\textrm{c}},
\end{equation}
 where $I_{\textrm{i1,j1}}$ and $I_{\textrm{i2,j2}}$ are the unsigned magnetic intensities (absolute values of $B_\textrm{LOS}$) of any two arbitrary pixels with different polarities, and $I_{\textrm{c}}$ corresponds to the absolute values of all pixels placed along the line joining the two pixels. For example, in the HMI cutout image of Figure 1, the two pixels (red asterisks in panels a and b) connect only if their magnetic intensities exceed the values of those pixels laid on the line (panel c). Note that the likelihood of a link must be examined only between nodes with opposite polarities (panel d). A physical approach to constructing a magnetic network must focus on connections between positive and negative regions. In the remainder of this paper, we show how the graph theory conveniently accomplishes identifying magnetic patches.

\section{The Network's Properties}\label{sec:properties}

Having the magnetic network constructed, the calculation of its parameters is required for further investigations. To this purpose, we first calculate the adjacency matrix that contains information on the graph's nodes (i.e., pixel locations) and edges (i.e., connectivity). Generally, for a magnetogram image of size $ m \times n $ pixels, there are $ N = m \times n $ nodes over which the connectivity must be checked. The size of the adjacency matrix for such a graph is $ N^2 $. Introducing a threshold for the background field could practically decrease the execution time as it removes some of the nodes and shrinks the adjacency matrix. Generally, thresholds higher than 12 G are appropriate \citep [see e.g.,][and the references therein]{shokri2022synchronization}.

For a simple undirected and unweighted graph, the adjacency matrix is a symmetric array with elements equal to either $1$ or $0$. These values indicate whether or not a connection is established between nodes. However, in directed networks, the matrix's elements could adopt either positive or negative signs as a representation of the entry or exhaust of the edges into the nodes. In the case of weighted networks, there are no limitations, and the adjacency matrix could have any value (rather than binaries) manifesting the importance of the established connections. 

We construct a directed and weighted graph to study the solar magnetic patches. We consider the incoming/outgoing magnetic intensities as the weight of connections. Specifically, if a connection is established between pixels $i$ and $j$, then $A_\textrm{i,j}$ equals to $B_\textrm{LOS}$ of pixel $i$, and $A_\textrm{j,i}$ equals to $B_\textrm{LOS}$ of pixel $j$. In case of no connectivity $A_\textrm{i,j}$ would be zero. Also, $A_\textrm{i, i}$ is considered to be zero, due to the nature of the magnetic network. The next step is to investigate the graph's properties (e.g., degree distribution and PageRank). 

The degree distribution specifies how many effective connections are established in a network by measuring the number of nodes' neighbors. By definition, the degree of the $i$-th node of a graph is:
\begin{equation}\label{degree}
	k_{\textrm{i}}=\sum_{\textrm{j=1}}^{N}A_{\textrm{i,j}},
\end{equation}
where $A$ is the adjacency matrix \citep{donges}.

Further to the degree distribution, we calculate the PageRank and assess its applicability in the recognition of magnetic patches. The PageRank, $r_{\textrm{i}},$ illustrates the importance (popularity) of a node based on the structure of links in a graph \citep{sheng}:
\begin{equation}\label{degree}
	r_{\textrm{i}}=\frac{1-d}{N}+d \sum_{\textrm{j} \in {N}}\frac{r_{\textrm{j}}}{k_{\textrm{j}}}.
\end{equation}
In this equation, the damping factor $d$ is a constant and could adopt any value between $0$ and $1$. But usually, it is considered to be $0.85$ \citep{BRIN1998107, zohre}.

In the next section, we discuss these properties in more detail and examine their usefulness in the detection of magnetic patches. We show that the map of degree distribution and PageRank could distinguish the borders of photospheric fluxes on the Sun's surface.

\begin{figure*}
\centering
		\includegraphics[width=8cm,height=7cm]{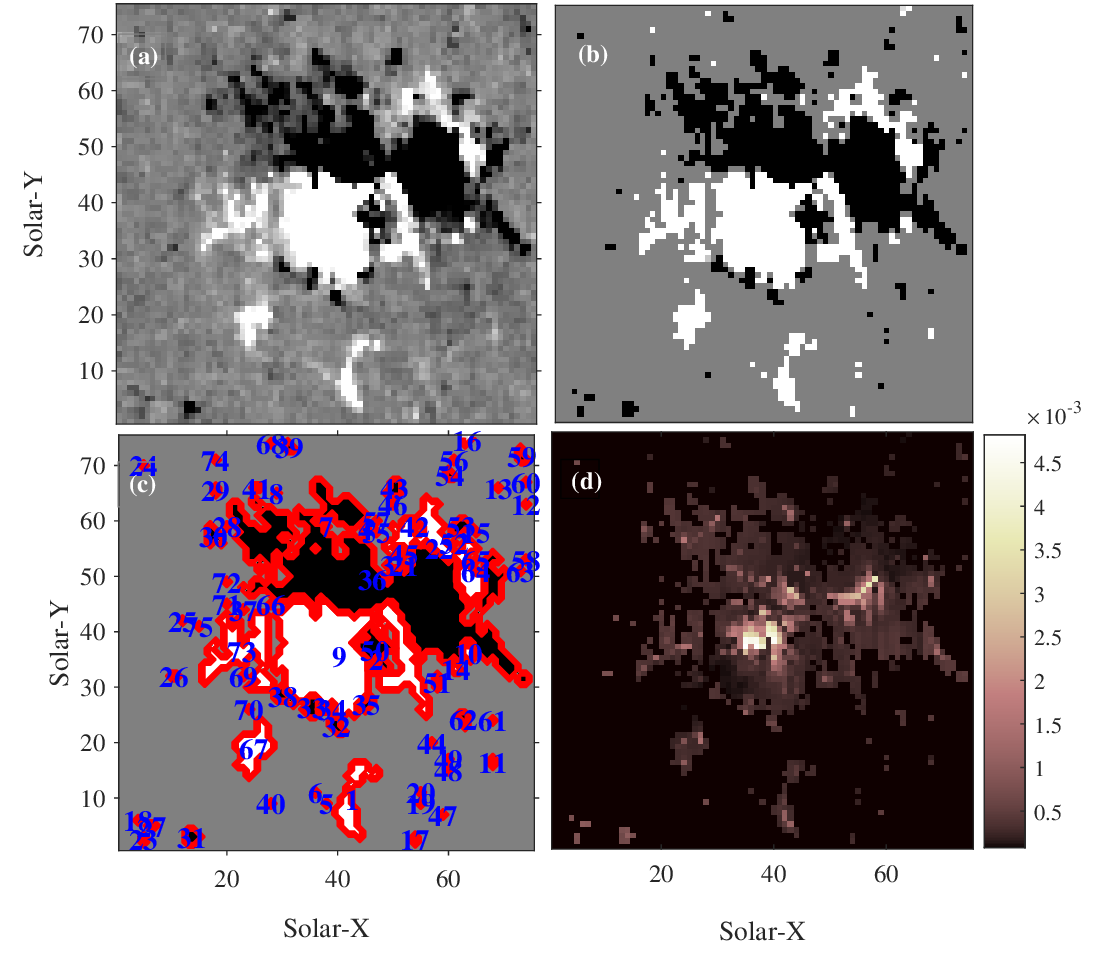}
		\caption{(a) The HMI cutout image of an AR recorded by the SDO at 23:58 on January 17, 2022. The selected window size is 75 $\times$ 75 pixels. The complex network is constructed considering a threshold of about 20 G. (b) The map of the degree distribution; (c) Identified magnetic patches; (d) The PageRank map.}
		\label{Fig3}
	\end{figure*} 

\section{Results and discussions}\label{results}
The solar magnetic activity could be investigated by detecting magnetic structures as they emerge, evolve, and annihilate on the surface. Following this objective, we introduced a new algorithm based on the complex network approach to identify the photospheric magnetic patches from HMI LOS images (the relevant MATLAB and Python packages are available on GitHub, see Data Availability Section). The applicability and efficiency of the developed algorithm are assessed through the examination of various data sets and the evaluation of the networks' properties (i.e., degree distribution and PageRank). These properties are found to profitably identify magnetic patches from their environment.
\begin{figure}
		\centering\includegraphics[width=8cm,height=7cm]{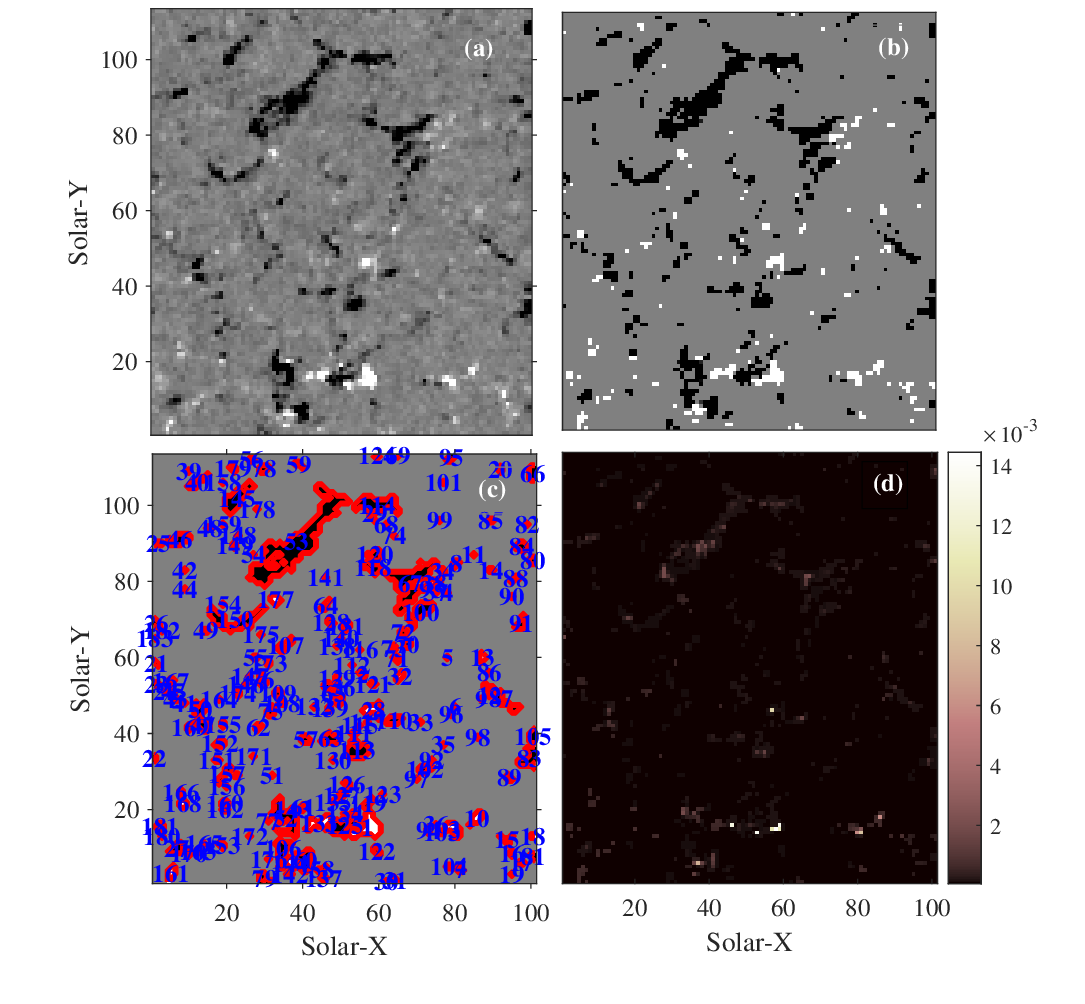}
		\caption{(a) The HMI cutout image of a quiet region recorded by the SDO at 23:59 on July 12, 2022. The selected window size is 113 $\times$ 101 pixels; The complex network is constructed considering a threshold of about 18 G. (b) The map
of the degree distribution; (c) Identified magnetic patches; (d) The PageRank map.}
		\label{Fig4}
\end{figure}	

Considering a minimum background threshold for $B_\textrm{LOS}$ could affect the execution time and the detection of small patches. Figure \ref{Fig2} shows an HMI cutout image of an AR (NOAA 2929), registered at 23:58 on January 17, 2022. The Figure presents the identified magnetic islands for various choices of threshold i.e., 8, 12, 16, 20, and 24 G. Seemingly, lower thresholds result in more intricate outcomes. Considering that the inbred HMI magnetogram noise level hinges on the instrument data product and the area of interest in the photosphere, different thresholds are applicable for various studies

 Figure 3 shows an HMI cutout image (panel a) and the map of the network’s degree distribution and PageRank (panels b,d respectively). Magnetic features are clearly visible on the map of the degree distribution. We acknowledge that the original maps have varying shades due to the wide range of plausible values for the degrees. However, a uniform color scheme is applied to all the degree distribution maps to render a better manifestation of borders. We obtain that higher magnetic fluxes result in higher values for PageRank. A similar analysis is performed on the HMI image of a quiet region recorded on July 12, 2021 (Figure \ref{Fig4}). Again, we observe that the visibility graph approach could detect the magnetic patches conveniently. 

Borders of the identified features could be determined by applying any arbitrary region-growing method. Here, we apply the Downhill algorithm and label the patches based on their connectivity, i.e., summation over all nodes' degrees within each region (Figures \ref{Fig3} and \ref{Fig4}, panel c). This algorithm is basically utilized on HMI images for the recognition of photospheric features, but here, we use it on both the distribution maps and HMI images to extract the borders and evaluate the complex network's performance, respectively. 

\begin{figure*}
  \centering
       \includegraphics[width=12cm,height=6.8cm]{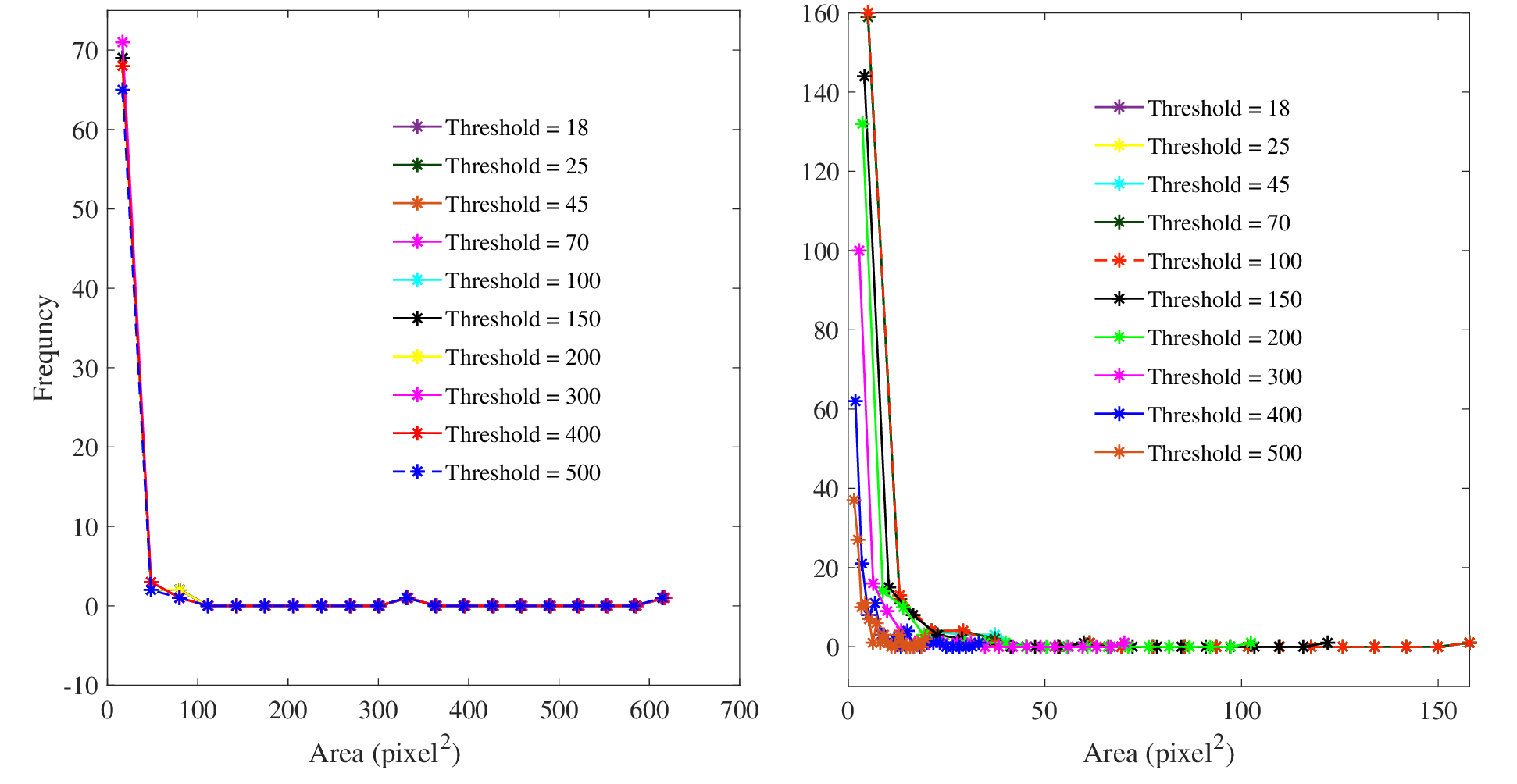}
					\caption{Patch sizes for various choices of threshold applying the Downhill algorithm: (a) in an AR; (b) in a quiet Sun. }
		\label{Fig5}
\end{figure*}

Given an initial threshold, Downhill divides the identified regions into smaller sub-regions. Figure \ref{Fig5} illustrates the impact of the threshold on the size of patches for two active and quiet regions, panels (a) and (b), respectively. The choice of the threshold manifests no significant effect on the patch sizes in an AR (up to 150 G). However, the threshold demonstrates a more profound influence over the quiet Sun, as higher thresholds provide less accurate determinations by neglecting small features. This could particularly affect statistical studies of magnetic patches and elaborate precautions need to be taken in this regard. 

\begin{figure*}
\centering
\includegraphics[width=12cm,height=6.8cm]{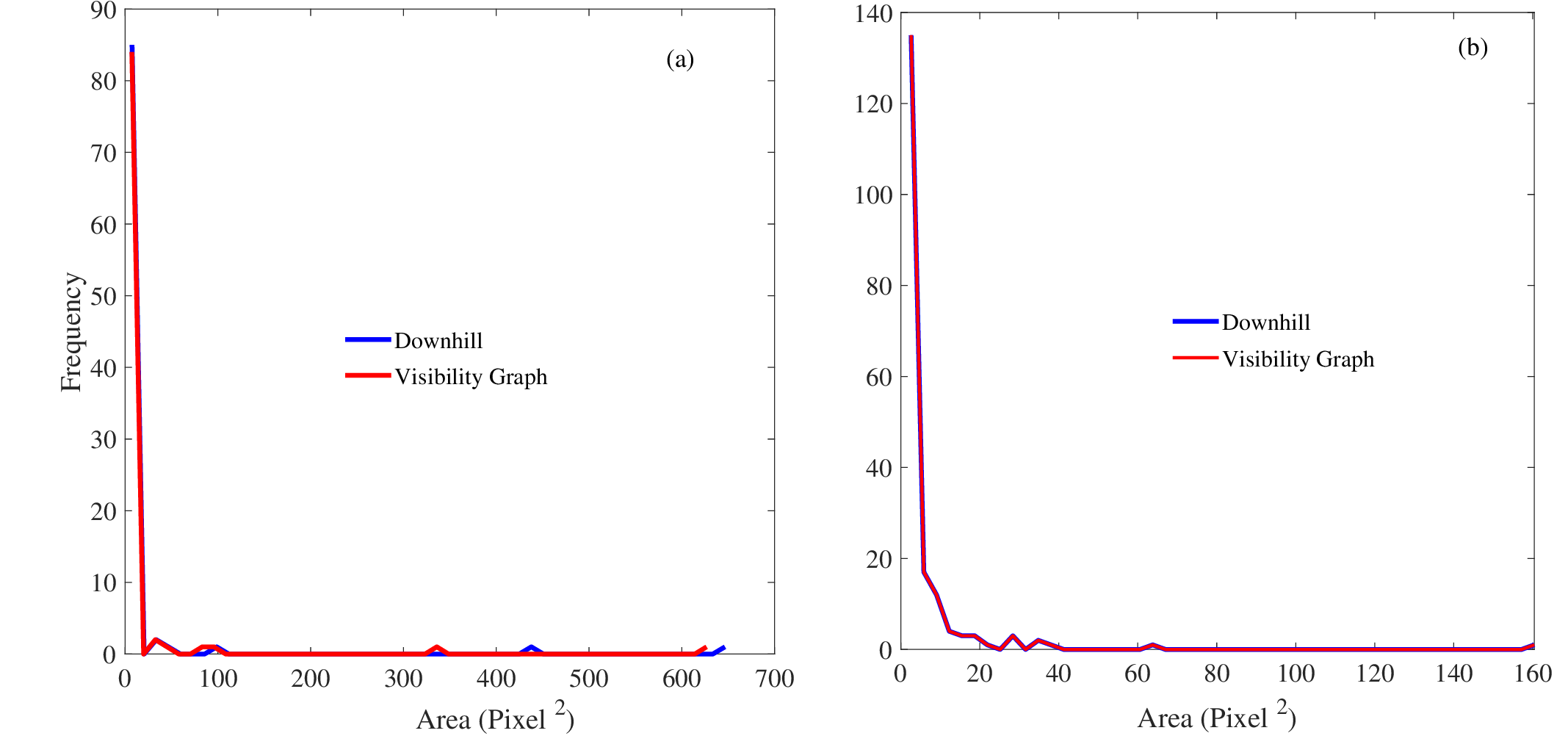}	
		\caption{ Frequency-size distribution of magnetic patches identified by the conventional Downhill technique (blue line) and complex network methods (red line) for an AR (a) and a quiet Sun (b).}
		\label{Fig6}
\end{figure*}
A comparison between the Downhill identification algorithm and the complex magnetic network approach is made. Figure \ref{Fig6} displays the frequency-size distribution of patches identified by the Downhill method applied to the network distribution maps (red line) and to the HMI images (blue line) for an AR and a quiet Sun at a given threshold. The results based on the network algorithm are well-matched with the downhill method.

\begin{figure}
    \centering
        \includegraphics[width=9cm,height=8cm]{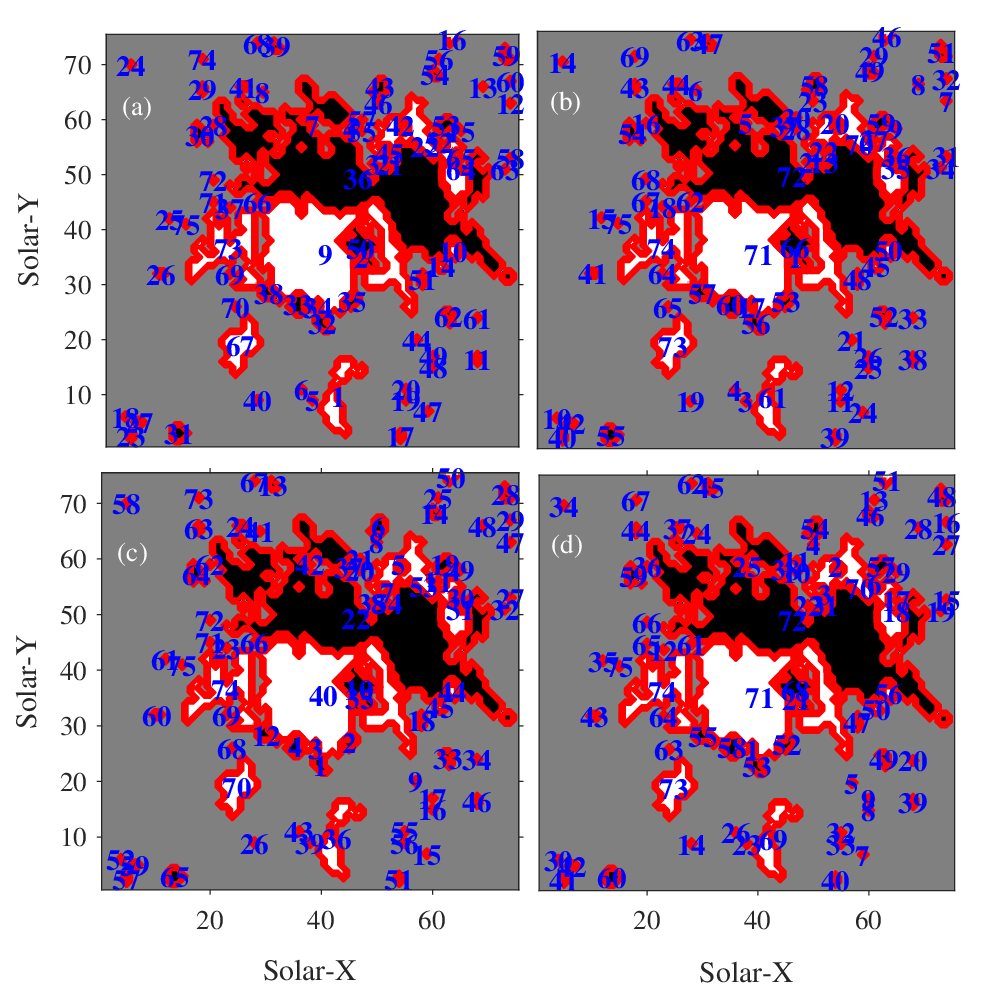}\\   
		\caption{The identified magnetic patches ranked by numbers according to: (a) the total degree of the nodes; (b) the average degree of the node for pixels  (divided by patch size); (c) the total magnetic field; (d) the average magnetic field for pixels (divided by patch size). }
		\label{Fig7}
	\end{figure}

The solar surface is covered by magnetic patches and the atmospheric structures/phenomena have their origins in these regions. Nonetheless, not all the patches share the same impact on this environment. Parameters such as size, lifetime, and magnetic field strength are likely to regulate the effectiveness (importance) of magnetic patches. Accordingly, in traditional algorithms, the key features were characterized based on these parameters. Alternatively, we propose the examination of magnetic connectivity. The complex network approach provides the ability to rank the magnetic patches based on their affinity (i.e., degree of nodes) and importance (i.e., PageRank) by searching for connections between opposite polarities. A technical differentiation exists between the present perspective and actual magnetic connections, i.e., field lines. However, there is a possibility of discovering commonalities between the two. Figure \ref{Fig7} presents the magnetic features ranked based on their overall degrees and total magnetic fields. Indeed, further investigations could provide a better understanding of this perspective and its efficiency in studies of atmospheric events. 

\begin{figure}
 \centering
	   \includegraphics[width=13.6cm,height=8cm]{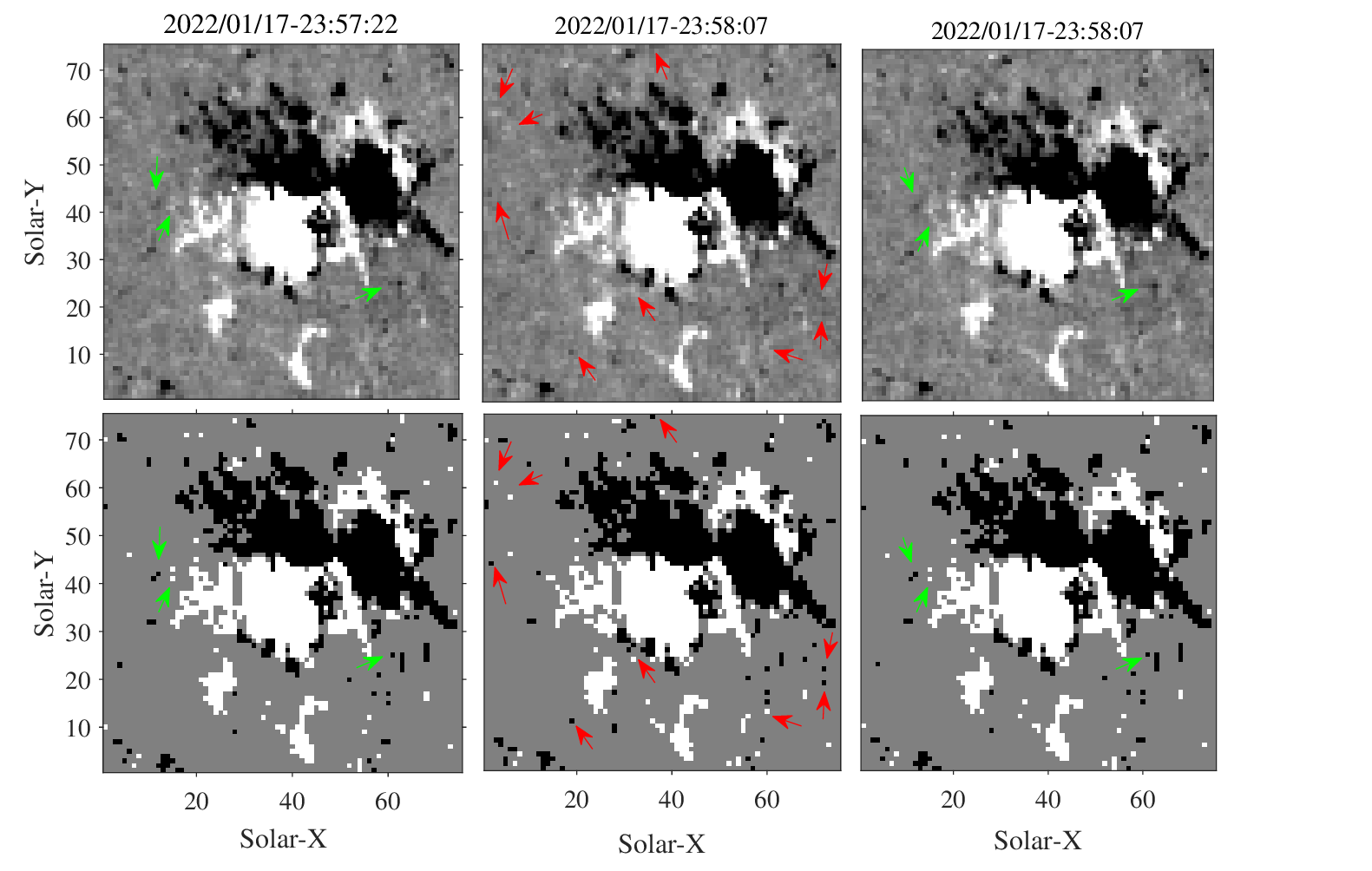}	
		\caption{Consecutive HMI magnetograms with the cadence of 45 s (top panels) together with the identified features based on the network approach (bottom panels). The red arrows  in the middle magnetograms point at the identified small-scale (one or two pixels) events that did not appear in the left or right panels. The patches with green arrows correspond to the small-scale events of the corner panels that were not detected in the middle frame.}
		\label{Fig8}
\end{figure}

Besides comparing with Downhill, we track the identified patches in consecutive frames to evaluate the efficiency of the network approach particularly in detecting the small-scale (one or two pixels) patches. An event is most likely non-noisian if it appears in two or more successive images. Figure \ref{Fig8} shows an example of three consecutive HMI magnetograms with a time cadence of 45 s versus their identified patches. Out of 74 identified small-scale features detected in the middle magnetogram, 9 features are not detected in the left or right magnetograms (shown with red arrows). Performing a similar analysis on more frames gives an average of 8\% false-positive error in the detection of small-scale features. Such features are either noise or concise lifetime patches that cannot be confirmed using the HMI data with 45 s cadences. Furthermore, we observe an average of 1\% false-negative errors. The false-negative error indicates the failure in the identification of small-scale patches within an image which appeared in the preceding and following magnetograms. These features are primarily non-noisian events.

One might think of the complex network approach analogous to any other flux-based identification algorithm. Such a misinterpretation might be inspired by the fact that the networks are constructed based on magnetic intensities. However, the measure of the correlation between the magnetic field and the degree of each node, $ \sim 0.44$, suggests a clear discrepancy. Furthermore, other approaches to identification usually confront some restrictions. These limitations mainly regard the size of magnetic patches, the number of frames required to acknowledge a feature, and the choice of the intensity threshold \citep{hagenaar1999dispersal, parnell2002nature, welsch2002magnetic, Colak, caballero}. The complex network approach successfully identifies small-scale patches (even those as small as one pixel). Yet, the optional choice of background intensity seems to play a major role in the accuracy of the identifications (Figure \ref{Fig2}). 

The dependency of the present method on the predefined threshold intrinsically differs from the previous algorithms. The susceptibility of other approaches to identification to noise and misidentifications mostly originates in the applied region-growing algorithm. On the other hand, the relatively unfavorable (noisy-like) identification of the visibility graph method at low thresholds corresponds to the small fragments of the photospheric magnetic carpet (Figures \ref{Fig1} d and \ref{Fig2}). In other words, the noisy-like features might be of interest depending on the intended study and various levels of accuracy are convenient.

We performed the analysis on several other data sets and obtained that the proposed visibility-graph-based algorithm serves as an efficient and reliable means of identification for solar magnetic patches. The routine remarkably succeeds in the fast recognition of magnetic features in cutout images. As the next step, we intend to investigate the temporal evolution of the small and large solar magnetic patches via this algorithm. We aim to discern noise from tiny features and track the sunspot regions.

\vspace{5mm}

\textbf{Data Availability:} The developed MATLAB and Python packages for Identifying Solar Magnetic Patches (ISMP) are available at \\
\url{https://github.com/Zahra-Tajik/ISMP.git}.

\vspace{5mm}

\textbf{Acknowledgments:} The authors gratefully thank the NASA/SDO science group and the JSOC central data repository for making data publicly available. We also acknowledge the use of the YAFTA algorithm.

\clearpage
\bibliographystyle{apalike}

\bibliography{bibtex.bib}

\begin{thebibliography}{}

\bibitem[Archontis and Syntelis, 2019]{archontis2019emergence}
Archontis, V. and Syntelis, P. (2019).
\newblock The emergence of magnetic flux and its role on the onset of solar
  dynamic events.
\newblock {\em Philosophical Transactions of the Royal Society A},
  377(2148):20180387.

\bibitem[Arish et~al., 2016]{mohsen}
Arish, S., Javaherian, M., Safari, H., and Amiri, A. (2016).
\newblock Extraction of active regions and coronal holes from euv images using
  the unsupervised segmentation method in the bayesian framework.
\newblock {\em Solar Physics}, 291(4):1209--1224.

\bibitem[Barra et~al., 2008]{barra}
Barra, V., Delouille, V., and Hochedez, J.-F. (2008).
\newblock Segmentation of extreme ultraviolet solar images via multichannel
  fuzzy clustering.
\newblock {\em Advances in Space Research}, 42(5):917--925.

\bibitem[Barra et~al., 2009]{barrafast}
Barra, V., Delouille, V., Kretzschmar, M., and Hochedez, J.-F. (2009).
\newblock Fast and robust segmentation of solar euv images: algorithm and
  results for solar cycle 23.
\newblock {\em Astronomy \& Astrophysics}, 505(1):361--371.

\bibitem[Bellot~Rubio and Orozco~Su{\'a}rez, 2019]{bellot2019quiet}
Bellot~Rubio, L. and Orozco~Su{\'a}rez, D. (2019).
\newblock Quiet sun magnetic fields: an observational view.
\newblock {\em Living Reviews in Solar Physics}, 16(1):1--124.

\bibitem[Benkhalil et~al., 2006]{benkha}
Benkhalil, A., Zharkova, V., Zharkov, S., and Ipson, S. (2006).
\newblock Active region detection and verification with the solar feature
  catalogue.
\newblock {\em Solar Physics}, 235(1):87--106.

\bibitem[Bo et~al., 2022]{bo2022automatic}
Bo, J., Lei, L., Sheng, Z., Shan-shan, Y., Shu-guang, Z., Yao, H., and Xiao-yu,
  L. (2022).
\newblock An automatic detection for solar active regions based on
  scale-invariant feature transform and clustering by fast search and find of
  density peaks.
\newblock {\em Chinese Astronomy and Astrophysics}, 46(3):264--276.

\bibitem[Boers et~al., 2015]{boers}
Boers, N., Donner, R.~V., Bookhagen, B., and Kurths, J. (2015).
\newblock Complex network analysis helps to identify impacts of the el ni{\~n}o
  southern oscillation on moisture divergence in south america.
\newblock {\em Climate Dynamics}, 45(3):619--632.

\bibitem[Brin and Page, 1998]{BRIN1998107}
Brin, S. and Page, L. (1998).
\newblock The anatomy of a large-scale hypertextual web search engine.
\newblock {\em Computer Networks and ISDN Systems}, 30(1):107--117.
\newblock Proceedings of the Seventh International World Wide Web Conference.

\bibitem[Caballero and Aranda, 2014]{caballero}
Caballero, C. and Aranda, M. (2014).
\newblock Automatic tracking of active regions and detection of solar flares in
  solar euv images.
\newblock {\em Solar Physics}, 289(5):1643--1661.

\bibitem[Cheung et~al., 2007]{cheung2007magnetic}
Cheung, M. C.~M., Schuessler, M., and Moreno-Insertis, F. (2007).
\newblock Magnetic flux emergence in granular convection: radiative mhd
  simulations and observational signatures.
\newblock {\em Astronomy \& Astrophysics}, 467(2):703--719.

\bibitem[Cho et~al., 2007]{cho2007magnetic}
Cho, K.-S., Lee, J., Gary, D., Moon, Y.-J., and Park, Y. (2007).
\newblock Magnetic field strength in the solar corona from type ii band
  splitting.
\newblock {\em The Astrophysical Journal}, 665(1):799.

\bibitem[Daei et~al., 2017]{Farhad}
Daei, F., Safari, H., and Dadashi, N. (2017).
\newblock Complex network for solar active regions.
\newblock {\em The Astrophysical Journal}, 845(1):36.

\bibitem[DeForest et~al., 2007]{deforest2007solar}
DeForest, C., Hagenaar, H., Lamb, D., Parnell, C., and Welsch, B. (2007).
\newblock Solar magnetic tracking. i. software comparison and recommended
  practices.
\newblock {\em The Astrophysical Journal}, 666(1):576.

\bibitem[DeRosa and Slater, 2013]{DeRosa}
DeRosa, M. and Slater, G. (2013).
\newblock Guide to sdo data analysis.
\newblock {\em Lockheed Martin Solar \& Astrophysics Laboratory, Palo Alto,
  CA}.

\bibitem[Donges et~al., 2009]{donges}
Donges, J.~F., Zou, Y., Marwan, N., and Kurths, J. (2009).
\newblock Complex networks in climate dynamics.
\newblock {\em The European Physical Journal Special Topics}, 174(1):157--179.

\bibitem[Farhang et~al., 2018]{Farhang2018}
Farhang, N., Safari, H., and Wheatland, M.~S. (2018).
\newblock Principle of minimum energy in magnetic reconnection in a
  self-organized critical model for solar flares.
\newblock {\em The Astrophysical Journal}, 859(1):41.

\bibitem[Farhang et~al., 2022]{Farhang2022}
Farhang, N., Shahbazi, F., and Safari, H. (2022).
\newblock Do cellular automaton avalanche models simulate the quasi-periodic
  pulsations of solar flares?
\newblock {\em The Astrophysical Journal}, 936(1):87.

\bibitem[Farhang et~al., 2019]{Farhang2019}
Farhang, N., Wheatland, M.~S., and Safari, H. (2019).
\newblock Energy balance in avalanche models for solar flares.
\newblock {\em The Astrophysical Journal Letters}, 883(1):L20.

\bibitem[Gheibi et~al., 2017]{gheibi}
Gheibi, A., Safari, H., and Javaherian, M. (2017).
\newblock The solar flare complex network.
\newblock {\em The Astrophysical Journal}, 847(2):115.

\bibitem[Hagenaar et~al., 1999]{hagenaar1999dispersal}
Hagenaar, H., Schrijver, C., Shine, R., et~al. (1999).
\newblock Dispersal of magnetic flux in the quiet solar photosphere.
\newblock {\em The Astrophysical Journal}, 511(2):932.

\bibitem[Harker, 2012]{harker}
Harker, B.~J. (2012).
\newblock Parameter-free automatic solar active region detection by hermite
  function decomposition.
\newblock {\em The Astrophysical Journal Supplement Series}, 203(1):7.

\bibitem[Higgins et~al., 2011]{Higgins}
Higgins, P.~A., Gallagher, P.~T., McAteer, R.~J., and Bloomfield, D.~S. (2011).
\newblock Solar magnetic feature detection and tracking for space weather
  monitoring.
\newblock {\em Advances in Space Research}, 47(12):2105--2117.

\bibitem[Kaki et~al., 2022]{kaki}
Kaki, B., Farhang, N., and Safari, H. (2022).
\newblock Evidence of self-organized criticality in time series by the
  horizontal visibility graph approach.
\newblock {\em Scientific Reports}, 12.

\bibitem[Kestener et~al., 2010]{kestener}
Kestener, P., Conlon, P., Khalil, A., Fennell, L., McAteer, R., Gallagher, P.,
  and Arneodo, A. (2010).
\newblock Characterising complexity in compound systems: Segmentation in
  wavelet-space.
\newblock {\em Astrophysical Journal}, 717:995--1005.

\bibitem[Kosovichev, 2009]{kosovichev2009photospheric}
Kosovichev, A. (2009).
\newblock Photospheric and subphotospheric dynamics of emerging magnetic flux.
\newblock {\em Space science reviews}, 144(1):175--195.

\bibitem[Lamb and Deforest, 2003]{lamb2003magnetic}
Lamb, D. and Deforest, C. (2003).
\newblock Magnetic element tracking and the small-scale solar dynamo.
\newblock In {\em AGU Fall Meeting Abstracts}, volume 2003, pages SH42B--0530.

\bibitem[McAteer et~al., 2005]{mcateer}
McAteer, R., Gallagher, P.~T., Ireland, J., and Young, C.~A. (2005).
\newblock Automated boundary-extraction and region-growing techniques applied
  to solar magnetograms.
\newblock {\em Solar Physics}, 228(1):55--66.

\bibitem[Miesch, 2005]{miesch2005large}
Miesch, M.~S. (2005).
\newblock Large-scale dynamics of the convection zone and tachocline.
\newblock {\em Living Reviews in Solar Physics}, 2(1):1--139.

\bibitem[Mohammadi et~al., 2021]{zohre}
Mohammadi, Z., Alipour, N., Safari, H., and Zamani, F. (2021).
\newblock Complex network for solar protons and correlations with flares.
\newblock {\em Journal of Geophysical Research: Space Physics},
  126(7):e2020JA028868.

\bibitem[Murray et~al., 2006]{murray20063d}
Murray, M., Hood, A., Moreno-Insertis, F., Galsgaard, K., and Archontis, V.
  (2006).
\newblock 3d simulations identifying the effects of varying the twist and field
  strength of an emerging flux tube.
\newblock {\em Astronomy \& Astrophysics}, 460(3):909--923.

\bibitem[Najafi et~al., 2020]{najafi}
Najafi, A., Darooneh, A.~H., Gheibi, A., and Farhang, N. (2020).
\newblock Solar flare modified complex network.
\newblock {\em The Astrophysical Journal}, 894(1):66.

\bibitem[N{\'o}brega-Siverio et~al., 2020]{nobrega2020nonequilibrium}
N{\'o}brega-Siverio, D., Moreno-Insertis, F., Martinez-Sykora, J., Carlsson,
  M., and Szydlarski, M. (2020).
\newblock Nonequilibrium ionization and ambipolar diffusion in solar magnetic
  flux emergence processes.
\newblock {\em Astronomy \& Astrophysics}, 633:A66.

\bibitem[Parker, 1955]{parker1955hydromagnetic}
Parker, E.~N. (1955).
\newblock Hydromagnetic dynamo models.
\newblock {\em The Astrophysical Journal}, 122:293.

\bibitem[Parnell, 2002]{parnell2002nature}
Parnell, C. (2002).
\newblock Nature of the magnetic carpet--i. distribution of magnetic fluxes.
\newblock {\em Monthly Notices of the Royal Astronomical Society},
  335(2):389--398.

\bibitem[Pesnell et~al., 2011]{Pesnell}
Pesnell, W.~D., Thompson, B.~J., and Chamberlin, P. (2011).
\newblock The solar dynamics observatory (sdo).
\newblock In {\em The solar dynamics observatory}, pages 3--15. Springer.

\bibitem[Priest, 2014]{priest2014magnetohydrodynamics}
Priest, E. (2014).
\newblock {\em Magnetohydrodynamics of the Sun}.
\newblock Cambridge University Press.

\bibitem[Qahwaji and Colak, 2005]{Colak}
Qahwaji, R. and Colak, T. (2005).
\newblock Automatic detection and verification of solar features.
\newblock {\em International Journal of Imaging Systems and Technology},
  15(4):199--210.

\bibitem[Quan et~al., 2021]{quan}
Quan, L., Xu, L., Li, L., Wang, H., and Huang, X. (2021).
\newblock Solar active region detection using deep learning.
\newblock {\em Electronics}, 10(18):2284.

\bibitem[Scherrer et~al., 2012]{scherrer}
Scherrer, P.~H., Schou, J., Bush, R., Kosovichev, A., Bogart, R., Hoeksema, J.,
  Liu, Y., Duvall, T., Zhao, J., Schrijver, C., et~al. (2012).
\newblock The helioseismic and magnetic imager (hmi) investigation for the
  solar dynamics observatory (sdo).
\newblock {\em Solar Physics}, 275(1):207--227.

\bibitem[Schmieder et~al., 2014]{schmieder2014magnetic}
Schmieder, B., Archontis, V., and Pariat, E. (2014).
\newblock Magnetic flux emergence along the solar cycle.
\newblock {\em Space Science Reviews}, 186(1):227--250.

\bibitem[Schou et~al., 2012]{schou2012design}
Schou, J., Scherrer, P.~H., Bush, R.~I., Wachter, R., Couvidat, S.,
  Rabello-Soares, M.~C., Bogart, R.~S., Hoeksema, J., Liu, Y., Duvall, T.,
  et~al. (2012).
\newblock Design and ground calibration of the helioseismic and magnetic imager
  (hmi) instrument on the solar dynamics observatory (sdo).
\newblock {\em Solar Physics}, 275(1):229--259.

\bibitem[Sheng et~al., 2020]{sheng}
Sheng, J., Zhu, J., Wang, Y., Wang, B., Hou, Z., et~al. (2020).
\newblock Identifying influential nodes of complex networks based on
  trust-value.
\newblock {\em Algorithms}, 13(11):280.

\bibitem[Shokri et~al., 2022]{shokri2022synchronization}
Shokri, Z., Alipour, N., Safari, H., Kayshap, P., Podladchikova, O., Nigro, G.,
  and Tripathi, D. (2022).
\newblock Synchronization of small-scale magnetic features, blinkers, and
  coronal bright points.
\newblock {\em The Astrophysical Journal}, 926(1):42.

\bibitem[Steinhaeuser et~al., 2010a]{Steinhaeuser}
Steinhaeuser, K., Chawla, N.~V., and Ganguly, A.~R. (2010a).
\newblock Complex networks in climate science: Progress, opportunities and
  challenges.
\newblock In {\em CIDU}, pages 16--26.

\bibitem[Steinhaeuser et~al., 2010b]{KARSTEN}
Steinhaeuser, K., Chawla, N.~V., and Ganguly, A.~R. (2010b).
\newblock Complex networks in climate science: Progress, opportunities and
  challenges.
\newblock In {\em CIDU}, pages 16--26.

\bibitem[Taran et~al., 2022]{taran2022complex}
Taran, S., Khodakarami, E., and Safari, H. (2022).
\newblock Complex network view to solar flare asymmetric activity.
\newblock {\em Advances in Space Research}, 70(8):2541--2550.

\bibitem[Tortosa-Andreu and Moreno-Insertis, 2009]{tortosa2009magnetic}
Tortosa-Andreu, A. and Moreno-Insertis, F. (2009).
\newblock Magnetic flux emergence into the solar photosphere and chromosphere.
\newblock {\em Astronomy \& Astrophysics}, 507(2):949--967.

\bibitem[Verbeeck et~al., 2013]{verbeeck}
Verbeeck, C., Higgins, P.~A., Colak, T., Watson, F.~T., Delouille, V., Mampaey,
  B., and Qahwaji, R. (2013).
\newblock A multi-wavelength analysis of active regions and sunspots by
  comparison of automatic detection algorithms.
\newblock {\em Solar Physics}, 283(1):67--95.

\bibitem[Watson et~al., 2009]{Watson}
Watson, F., Fletcher, L., Dalla, S., and Marshall, S. (2009).
\newblock Modelling the longitudinal asymmetry in sunspot emergence: the role
  of the wilson depression.
\newblock {\em Solar Physics}, 260(1):5--19.

\bibitem[Welsch and Longcope, 2002]{welsch2002magnetic}
Welsch, B. and Longcope, D. (2002).
\newblock Magnetic helicity injection by horizontal flows in the quiet sun: Ii.
  self helicity flux.
\newblock In {\em AGU Fall Meeting Abstracts}, volume 2002, pages SH52A--0455.

\bibitem[Welsch and Longcope, 2003]{Welsch2003}
Welsch, B. and Longcope, D. (2003).
\newblock Magnetic helicity injection by horizontal flows in the quiet sun. i.
  mutual-helicity flux.
\newblock {\em Astrophysical Journal - ASTROPHYS J}, 588:620.

\bibitem[Wiehr et~al., 2004]{wiehr2004brightness}
Wiehr, E., Bovelet, B., and Hirzberger, J. (2004).
\newblock Brightness and size of small-scale solar magnetic flux
  concentrations.
\newblock {\em Astronomy \& Astrophysics}, 422(3):L63--L66.

\bibitem[Zhang et~al., 2010]{zhang}
Zhang, J., Wang, Y., and Liu, Y. (2010).
\newblock Statistical properties of solar active regions obtained from an
  automatic detection system and the computational biases.
\newblock {\em The Astrophysical Journal}, 723(2):1006.

\bibitem[Zharkov et~al., 2005]{zharkov2005statistical}
Zharkov, S., Zharkova, V., and Ipson, S. (2005).
\newblock Statistical properties of sunspots in 1996--2004: I. detection,
  north--south asymmetry and area distribution.
\newblock {\em Solar Physics}, 228(1):377--397.

\bibitem[Zwaan, 1985]{zwaan1985emergence}
Zwaan, C. (1985).
\newblock The emergence of magnetic flux.
\newblock {\em Solar Physics}, 100(1):397--414.

\end{thebibliography}


\end{document}